\documentclass{moriond}

\def\Journal#1#2#3#4{{#1} {\bf #2}, #3 (#4)}

\def\NIMA{{\em Nucl. Instrum. Methods} A}

\def\PRD{{\em Phys. Rev.} D}

\def\EPJC{{\em Eur. Phys. J.} C}
\def\RMP{\em Rev. Mod. Phys.}

\def\be{\begin{equation}}
\def\ee{\end{equation}}
\def\bea{\begin{eqnarray}}
\def\eea{\end{eqnarray}}

\newcommand{\Photo}{}

\begin{document}
\vspace*{4cm}
\title{The Status and Prospects of the Muon $g-2$ Experiment at Fermilab}

\author{ A.T. FIENBERG }

\address{Department of Physics, University of Washington, Seattle}

\maketitle\abstracts{
The E989 Muon $g-2$ Experiment at Fermilab aims to measure the muon magnetic anomaly, $a_\mu$, to an unprecedented precision of 140 parts per billion (ppb), representing a four-fold improvement over the current best measurement, achieved at Brookhaven National Lab. There stands a greater than 3 standard deviations discrepancy between the Brookhaven measurement of $a_\mu$ and the theoretical value predicted using the Standard Model. The Fermilab experiment seeks to either resolve or confirm this discrepancy, which is suggestive of new physics interactions. To achieve the E989 target precision, the anomalous precession frequency of muons in a magnetic storage ring must be determined with a systematic uncertainty below 70\,ppb, and the average magnetic field experienced by these stored muons must be known equally well. The muon anomalous precession frequency is imprinted on the time-dependent energy distribution of decay positrons observed by 24 electromagnetic calorimeters. A suite of pulsed NMR probes continually monitors the magnetic field. This document presents the current status of the Fermilab experiment while emphasizing the ongoing analysis of the 2018 Run 1 dataset and the systematic effects that complicate it.}

\section{Motivation}

The magnetic dipole moment, $\vec{\mu}$, of a subatomic particle can be expressed in terms of its dimensionless $g$ factor as follows:
\begin{equation}
    \vec{\mu} = \pm g \frac{e}{2m} \vec{S},
\end{equation}
where $e$ is the elementary charge, $m$ is the particle's mass, and $S$ is its spin. 
The Dirac equation and tree-level quantum electrodynamics (QED) predict that $g=2$ for a structureless spin-1/2 particle such as the electron or the muon.
Loop effects yield observable adjustments to the Dirac equation's prediction. These adjustments motivate the definition of the magnetic anomaly, $a$:
\begin{equation}
    a \equiv \frac{g-2}{2}.
\end{equation}
While at tree level $a_e=a_\mu=0$, loop corrections to the electron and muon anomalies are significantly and measurably different. In general, $a_\mu$ receives larger contributions from virtual heavy particles than does $a_e$ because of the muon's larger mass. This document and the experiment it describes pertain to the muon magnetic anomaly, $a_\mu$.

The Standard Model (SM) provides a testable prediction of $a_\mu$, with QED accounting for the overwhelming majority of the predicted value.
Beyond QED, hadronic and electroweak (EW) effects produce 60\,parts-per-million (ppm) and 1\,ppm contributions, respectively. Recent comprehensive SM evaluations of $a_\mu$, such as those by Keshavarzi {\it et al}~\cite{knt} and Davier {\it et al},~\cite{dhmz} have combined uncertainties of 300--400\,parts-per-billion (ppb), and these uncertainties are dominated by nonperturbative hadronic interactions. Hadronic corrections are categorized by diagram topology into hadronic vacuum polarization (HVP) diagrams and hadronic light-by-light scattering (HLbL) diagrams. Currently, the HVP diagrams are evaluated using dispersive approaches
and the HLbL diagrams are evaluated using hadronic models. There is active research in the theory community toward reducing the SM hadronic uncertainties through the refinement of traditional techniques and the development of novel approaches.

The SM prediction of $a_\mu$ can be confronted with experiment. A measurement of $a_\mu$ that differed significantly from the SM prediction would be clear evidence of new phenomena. 
Conversely, confirmation of the SM prediction within the combined experimental and theoretical uncertainties would place severe constraints on proposed new physics models. Thus, a precision measurement of $a_\mu$ is a new physics search and SM test that is complementary to concurrent high-energy approaches.

Numerous experiments to measure $a_\mu$ have been conducted in the past decades. The E821 Experiment at Brookhaven National Lab (BNL) achieved a 540\,ppb measurement, which is sensitive to all categories of SM effects.~\cite{821} The BNL result differs from recent theoretical evaluations by 3.5--3.7$\sigma$, hinting at but not guaranteeing the presence of detectable new physics interactions. Despite careful scrutiny of both the experiment and the theory, the muon $g-2$ discrepancy observed at BNL has only increased in significance since its original publication. 
To elucidate the nature of the discrepancy, an improved measurement is warranted.

The E989 Experiment is designed to repeat the BNL measurement with a target uncertainty of 140\,ppb divided evenly between statistics and systematics.~\cite{TDR} E989 requires approximately 20 times the number of muons that were in the combined BNL dataset. Together with expected improvements in the theoretical prediction, E989 seeks to either resolve the BNL discrepancy or confirm it with a significance greater than 5$\sigma$.

\section{Experimental Technique}

When placed in a highly uniform dipole magnetic field---and within small, known corrections---the rate of change of the angle between a muon's spin and its momentum is directly proportional to $a_\mu$. This rate of change is called the anomalous precession frequency, $\omega_a$, and in the case where motion is entirely perpendicular to a perfectly uniform field it is related to $a_\mu$ by
\begin{equation}
    \label{eqn:oma}
    \omega_a = a_\mu \frac{eB}{m_\mu}.
\end{equation}

The precision of an $a_\mu$ value determined using Eq.~\ref{eqn:oma} is limited by whichever of the anomalous precession frequency, $\omega_a$, the magnetic field magnitude, $B$, the elementary charge, $e$, or the muon mass, $m_\mu$, is least precisely known. Introducing the muon-distribution-weighted average proton Larmor precession frequency in the storage ring's field, $\tilde{\omega_p}$, the proton magnetic moment $\mu_p$, the electron $g$ factor, $g_e$, the electron mass, $m_e$, and the electron magnetic moment, $\mu_e$, the above equation can be rearranged into the form
\begin{equation}
    \label{exp:eqn:amu}
    a_\mu = \frac{g_e}{2}\frac{\omega_a}{\tilde{\omega_p}}\frac{m_\mu}{m_e}\frac{\mu_p}{\mu_e}.
\end{equation}
The ratio $\omega_a / \tilde{\omega_p}$ 
will be measured in E989 and then combined with the quantities $g_e/2$, $m_\mu/m_e$, and $\mu_p/\mu_e$---known to 0.26\,parts-per-trillion, 22\,ppb, and 3\,ppb, respectively---to determine $a_\mu$. (Quoted uncertainties are CODATA's recommended values.)~\cite{codata}
The proxy for the storage ring's magnetic field, $\tilde{\omega_p}$, is measured using a suite of pulsed proton NMR probes: fixed probes that constantly monitor the magnetic field through time, probes mounted on a mobile trolley that intermittently measure the magnetic field in the muon storage region when beam is not present, an absolute calibration probe, and a probe that can be inserted or retracted from the storage ring to transfer the calibration from the absolute probe to the trolley probes. 

The quantity $\omega_a$ is measured solely through observing the energy spectrum of positrons produced by the decay of polarized $\mu^+$ in the storage ring. This is possible because of parity violation in the weak decay of the muon: in the rest frame of a positive muon, decay positrons are preferentially emitted in the direction of the muon spin. In the laboratory frame, positrons emitted in the direction of the muon momentum receive the largest possible Lorentz boost and are observed with higher energies than positrons emitted in other directions. Therefore, as the muon spins rotate relative to their momenta, the decay positron energy distribution in the laboratory frame changes. By this mechanism, the observed positron energy spectrum undergoes periodic modulation at $\omega_a$. 

Highly polarized 3.1\,GeV/$c$ muon bunches are delivered to the storage ring in a series of injections, or fills, at an average rate of 11.4\,Hz. The beam is injected with a radial offset relative to the design orbit, which is then corrected by firing a pulsed electromagnetic kicker once during the injected beam's first orbit. Electrostatic quadrupoles focus the beam vertically. Approximately 10,000 muons are stored per fill, and the minimum time separation between fills is 10\,ms.
Following each injection, muon decays are observed for approximately 700\,$\mu$s by 24 electromagnetic calorimeters that are evenly spaced in azimuth around the inside of the storage ring. In-vacuum straw tracking detectors are positioned in front of two of the 24 calorimeters to detect decay positrons in transit to those calorimeters, reconstruct their tracks, and extrapolate their trajectories back to the original decay vertices. In this way, the tracking detectors provide knowledge of the muon beam's physical position in the storage region as a function of time. This information is critical for understanding the stored beam's dynamics and the associated systematic uncertainties present in the $\omega_a$ measurement.

\section{Run 1 Analysis Status}

E989 completed its first commissioning run with beam in the spring of 2017 and its first physics run, Run 1, in the spring of 2018. Before data quality cuts, the final Run 1 dataset contains nearly twice as many observed muon decays as were present in the combined BNL dataset and is sufficient to reach an $a_\mu$ uncertainty of approximately 400\,ppb.
The collaboration is focused both on completing the Run 1 $a_\mu$ analysis in a timely fashion and on operating the experiment for Run 2. 

The Run 1 analysis efforts were initially limited to a subdataset, called the 60-Hour Dataset, sufficient for a 1.3\,ppm $\omega_a$ measurement. To ensure the correctness and internal consistency of the $\omega_a$ analysis, six separate teams worked in parallel to extract $\omega_a$. Across these six teams there were three separate treatments of the raw data (reconstructions). Each team independently developed the corrections necessary to remove certain undesirable effects from the data and determined the residual systematic uncertainties inherent in these corrections. 

A robust method for extracting $\omega_a$ from the calorimeter data is to set a fixed energy threshold and then count the number of decay positrons observed above that threshold. As the positron energy spectrum oscillates at $\omega_a$, so too does the probability that a decay positron will be emitted with an energy above any given nonzero threshold. Thus, in an idealized case, the times relative to the beam injection at which decay positrons above a certain energy threshold will be detected should be well described by the function
\begin{equation}
    \label{exp:eqn:fiveparam}
    f(t) = Ne^{-t/\tau}\left[1 + A\cos(\omega_a t - \phi)\right].
\end{equation}
The asymmetry, $A$, and normalization, $N$, and the initial phase, $\phi$, will depend on the chosen energy threshold. One finds that the statistical uncertainty achieved using this technique is minimized with an energy threshold of 1.7\,GeV. 

In reality, Eq.~\ref{exp:eqn:fiveparam} does not adequately describe the collected data. Corrections are necessary both for the dynamics of the stored beam and for the nonideal response of the calorimeters. The salient beam dynamics effect is the physical oscillation of the beam position in the storage region. As the muon beam is injected with a radial offset and with a range of momenta, no applied kick will place all injected muons onto their ideal orbits. Each individual muon oscillates about its ideal orbit, and the aggregate effect of all these oscillations is a coherent oscillation of the moments of the stored beam's vertical and radial distributions. These oscillations---which occur at known, calculable frequencies---appear in the calorimeter data through acceptance effects. The signals are quite strong and must be addressed by the fit model. Other beam dynamics effects that must be accounted for include stored muon losses, and corrections to Eq.~\ref{eqn:oma} arising from the electrostatic quadrupoles and from beam motion parallel to the magnetic field.

The primary detector-based effects of concern are pileup, or the calorimeter's inability to resolve arbitrarily close pulse pairs, and calorimeter gain changes. Pileup and gain can both bias $\omega_a$ through the energy dependence of the positron drift time. As evident in Eq.~\ref{exp:eqn:fiveparam}, a time shift is equivalent to a different phase, $\phi$. Time-dependent misinterpretation of the detected positron energies imparts to the observed phase a time dependence that is indistinguishable from a shifted precession frequency. In E989, gain changes are measured in situ with a laser calibration system~\cite{Anastasi} and removed in software. The differing reconstruction procedures employed by the six analysis teams have intrinsically different pileup behaviors, and each analysis team developed its own correction procedure to account for the pileup remaining after reconstruction.

The $\omega_a$ analyses are blinded; all fits are relative to a secret reference frequency. In February 2019, the six $\omega_a$ analysis teams compared their 60-Hour Dataset results using a common reference frequency---a relative unblinding. The numbers were in excellent agreement, and the groups produced similar estimates of the dataset's systematic uncertainties. This agreement validated the independent approaches taken by the different teams and demonstrated the collaboration's preparedness to analyze Run 1 in its entirety.

\section{Prospects}

Analysis efforts have moved beyond the 60-Hour Dataset. A comparison between analysis teams of blinded $\omega_a$ values extracted from the majority of the Run 1 data is planned for early summer 2019. Depending on the outcome of this comparison, final internal reviews and systematic uncertainty assessments will occur throughout the summer and a combined Run 1 $a_\mu$ result will be announced. Run 2 data collection is expected to continue until early summer 2019. Run 3 will occur in 2019-2020. 

\section*{Acknowledgments}

This work was supported in part by Fermilab and the US DOE, Office of Science, Office of Nuclear Physics under award number DE-FG02-97ER41020.

\end{document}